\documentclass[proceedings, preprint]{rmaa}

\usepackage{paralist}

\usepackage{psfrag,color}

\SetYear{2008}
\SetConfTitle{Magnetic Fields in the Universe II}

\title{Non-thermal emission from Galaxy Clusters and future observations
with the FERMI gamma-ray telescope and LOFAR}

\author{Gianfranco Brunetti\altaffilmark{1}}

\altaffiltext{1}{INAF- Istituto di Radioastronomia,
 via P. Gobetti, 101, Bologna, I-40129, Italy
 (brunetti@ira.inaf.it).}

\shortauthor{Gianfranco Brunetti}
\shorttitle{Particle acceleration in galaxy clusters}

\listofauthors{G. Bruneti}
\indexauthor{Brunetti, G.}

\abstract{FERMI (formely GLAST) and LOFAR will shortly provide crucial 
information on the
non-thermal components (relativistic particles and magnetic field)
in galaxy clusters. After discussing observational facts
that already put constraints on the properties and origin
of non-thermal components, I will report on the emission spectrum from
galaxy clusters as expected in the context of general calculations in which
relativistic particles (protons and secondary electrons due to
proton-proton collisions) interact with MHD turbulence generated in
the cluster volume during cluster-cluster mergers.
In this scenario (known as re-acceleration scenario) diffuse cluster-scale
radio emission is produced in massive clusters during merging events,
while gamma ray emission, at some level,
is expected to be common in clusters.
Expectations of interest for LOFAR and FERMI are also briefly
discussed.}

\resumen{}

\addkeyword{Turbulence}
\addkeyword{Acceleration of particles}
\addkeyword{Radiation mechanisms: non-thermal}
\addkeyword{Galaxies: clusters: general}

\begin{document}
\maketitle

\section{Introduction}
\label{sec:intro}

Clusters of galaxies are the largest gravitationally bound
objects in the present universe, containing
$\approx 10^{15}$ M$_{\odot}$ of hot ($10^8$ K) gas, galaxies and
dark matter.  The thermal gas, that is the dominant component in the
Inter-Galactic-Medium (IGM), is mixed with non-thermal components
such as magnetic fields and relativistic particles, as proved
by radio observations.
Non-thermal components play a key role by controlling
transport processes in the IGM (e.g. Narayan \& Medvedev 2001;
Lazarian 2006a) and are sources of additional pressure
(e.g. Ryu et al. 2003), thus their origin and evolution are important
ingredients to understand the physics of the IGM.

The bulk of present information on the non-thermal components comes
from radio telescopes that discovered an increasing number of
Mpc-sized diffuse radio sources in a fraction of galaxy clusters
(e.g. Feretti 2003; Ferrari et al. 2008). 
Cluster mergers are the most energetic events in the universe and are 
believed to power the mechanisms responsible for the origin of the 
non-thermal components in galaxy clusters. 
A fraction of the energy dissipated during these
mergers is expected to be channelled into the amplification of the
magnetic fields (e.g. Dolag et al. 2002; Subramanian et al. 2006)
and into the acceleration of particles via shocks and turbulence
that lead to a complex population of primary electrons and protons
in the IGM (e.g. Ensslin et al. 1998; Sarazin 1999; Blasi 2001;
Brunetti et al. 2001, 2004; Petrosian 2001; Miniati et al. 2001; Ryu
et al. 2003; Dolag 2006; Brunetti \& Lazarian 2007; Pfrommer 2008).
Theoretically relativistic protons are expected to be the dominant
non-thermal particles component since they have long
life-times and remain confined within galaxy clusters for a Hubble
time (e.g. Blasi et al. 2007 and ref. therein). Confinement enhances
the probability to have p-p collisions that in turns give gamma ray
emission via decay of $\pi^o$ produced during these
collisions and inject secondary particles that emit 
synchrotron and inverse Compton (IC) radiation whose intensity
depends on the energy density of protons. 
Only upper limits to the gamma ray emission from galaxy clusters have 
been obtained so far (Reimer et al. 2003), however the FERMI Gamma-ray
telescope\footnote{http://fermi.gsfc.nasa.gov/} (formely GLAST) 
will shortly allow a step
forward having a chance to obtain first detections of galaxy clusters 
or to put stringent constraints to the energy density of 
the relativistic protons.

The IGM is expected to be turbulent at some level and MHD turbulence
may re-accelerate both primary and secondary particles during
cluster mergers via second order Fermi mechanisms.
Turbulence is naturally generated in cluster mergers (Roettiger et al.
1999; Ricker \& Sarazin 2001; Dolag et al. 2005; Vazza et al. 2006;
Iapichino \& Niemeyer 2008) and the resulting
particle re-acceleration process should enhance the synchrotron
and IC emission by orders of magnitude
(e.g. Brunetti 2004; Petrosian \& Bykov 2008).
In a few years the Low Frequency Array (LOFAR) and the Long Wavelength Array
(LWA) will observe galaxy clusters at low radio frequencies (40-80 MHz and 
120-240 MHz in the case of LOFAR\footnote{http://www.lofar.org/} and 
10-88 MHz in the case of
LWA\footnote{http://lwa.nrl.navy.mil/}) catching 
the bulk of their synchrotron cluster-scale emission and
testing the different scenarios proposed for the origin of
the non-thermal particles.
Finally, the emerging pool of future hard X-ray telescopes
(e.g. NuSTAR, Simbol-X, Next) should provide crucial constraints
on the level of IC emission from clusters and on the strength of the
magnetic field in the IGM.

Facts from radio observations suggest that MHD turbulence
may play an important role in the acceleration of the electrons
responsible for the cluster-scale radio emission.
After briefly discussing observations (Sect.2),
in Sect.3 we will focus on re-acceleration models showing
expectations from the radio band to the gamma rays.

\section{Giant Radio Halos}
\label{sec:halos}

\subsection{Origin of the relativistic electrons in the IGM}
\label{sec:origin}

The most prominent examples of diffuse non-thermal sources in galaxy
clusters are the giant Radio Halos. These are low surface
brightness, Mpc-scale diffuse sources found at the centre of a
fraction of massive and merging galaxy clusters and are due to
synchrotron radiation from relativistic electrons diffusing in
$\approx \mu$G magnetic fields frozen in the IGM (Feretti 2003;
Ferrari et al. 2008). The origin of Radio Halos is still unclear, 
the starting point is that the timescale necessary for the emitting
electrons to diffuse over Radio Halo size-scales, of the order of a
Hubble time, is much longer than the electrons' radiative lifetime,
$\approx 10^8$ years (Jaffe 1977). 
This implies the existence of mechanisms of either {\it in situ} particle 
acceleration or injection into the IGM. 
Understanding the physics of 
these mechanisms is crucial to model the non-thermal components 
in galaxy clusters and their evolution.

Two main possibilities have been proposed to explain Radio Halos :
{\it i)} the so-called {\it re-acceleration} models, whereby
relativistic electrons injected in the IGM are
re-energized {\it in situ} by various mechanisms associated with the
turbulence in massive merger events (e.g. Brunetti et al. 2001;
Petrosian 2001; Fujita et al. 2003),
and {\it ii)} the {\it secondary electron} models,
whereby the relativistic electrons are secondary products of proton-proton
collisions between relativistic and thermal protons in the IGM
(e.g. Dennison 1980; Blasi \& Colafrancesco 1999; Dolag \& Ensslin 2000).

Extended and fairly regular radio emission is expected in the
case of a secondary origin of the emitting electrons, since the
parent primary protons can diffuse on large scales.
Still several properties of Radio Halos cannot be simply
reconciled with this scenario. Two ``{\it historical}''
points are :

\begin{itemize}

\item
Since all clusters have suffered mergers
(hierarchical scenario) and relativistic protons are mostly confined
within clusters, extended radio emission should be basically observed in
almost all clusters. On the other hand, Radio Halos are not common and,
although a fairly large number of clusters has an adequate radio follow up,
they are presently detected only in a fraction
of massive and merging clusters (Giovannini et al. 1999; Buote 2001;
Cassano et al. 2008; Venturi et al. 2008; Sect.\ref{sec:radio}).

\item
Since the spectrum of relativistic protons (and of
secondary electrons) is expected to be a simple power law
over a large range of particle-momentum, the synchrotron
spectrum of Radio Halos should be a power law.
On the other hand (although the spectral shape of Radio Halos is
still poorly known) the spectrum of the best studied Radio Halo, in
the Coma cluster, shows a cut off at GHz frequencies implying a 
corresponding cut off in the spectrum of the emitting electrons at 
$\approx$ GeV energy that was interpreted in favour of {\it in situ}
acceleration models (Schlickeiser et al. 1987; Thierbach et al. 2003).

\end{itemize}

\subsection{Inputs from new radio observations}
\label{sec:radio}

The two models in Sect.\ref{sec:origin} have different basic
expectations in terms of statistical properties of Radio Halos and
of their evolution. In particular, as already pointed out, in the
context of {\it secondary electron} models, Radio Halos should be
common and very long--living phenomena,  on the other hand, in the
context of {\it re-acceleration} models, due to the finite
dissipation time-scale of the turbulence in the IGM, they should be
{\it transient phenomena} with life--time $\approx$1 Gyr (or less) .

\begin{figure*}[!t]
\includegraphics[width=\columnwidth]{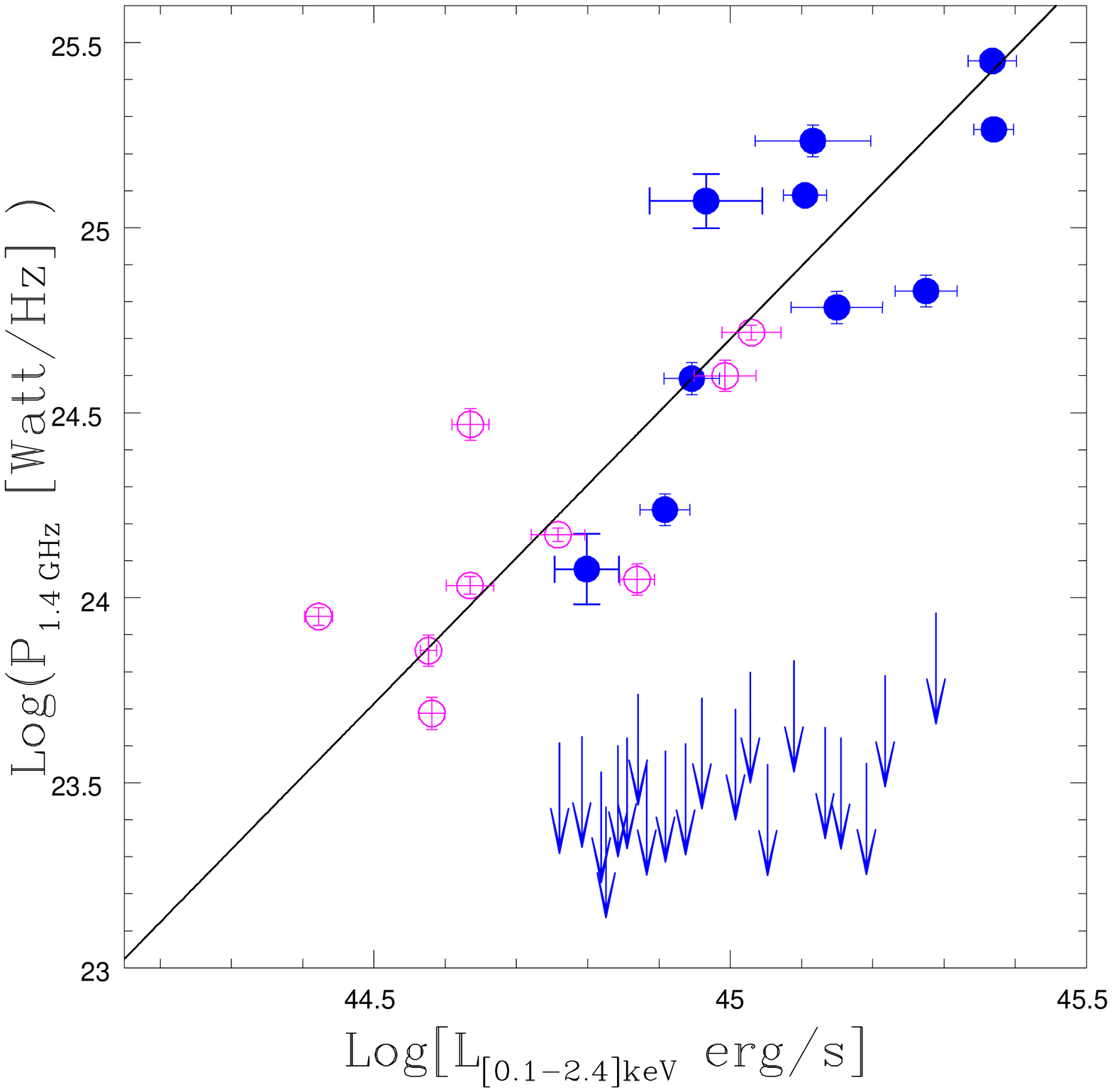}%
\hspace*{\columnsep}%
\includegraphics[width=\columnwidth]{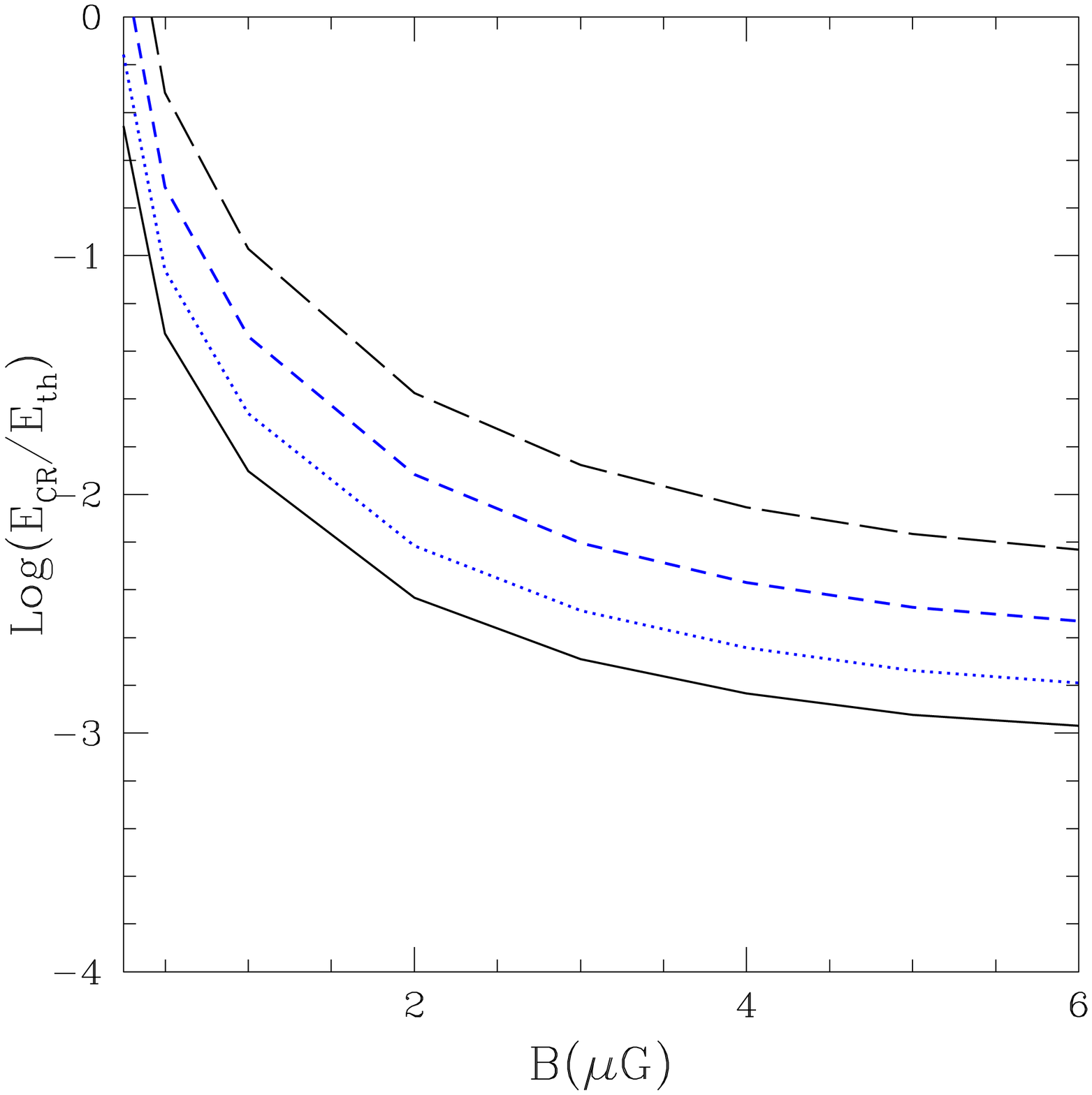}
\caption{{\bf Left Panel}: distribution of galaxy clusters in the
$P_{1.4}$--$L_x$ plane. Filled dots are galaxy clusters at
z$\geq$0.2 (from GMRT-sample and from the literature), empty dots
are clusters at lower redshift reported to highlight the
$P_{1.4}$--$L_x$ correlation (solid line, from Cassano et al. 2006).
Upper limits (represented by the arrows) are GMRT clusters with no
hint of cluster-scale emission and their distribution should be
compared with that of clusters at similar redshift (filled dots).
{\bf Right Panel}: upper limits (curves) to the ratio between the
energy density of relativistic protons (p $\geq 0.01$ mc) and that
of the IGM calculated for $\delta=$2.1, 2.3, 2.5, 2.7 (from bottom
to top) assuming a proton spectrum $N(p) \propto p^{-\delta}$ (see
Brunetti et al. 2007 for details).} \label{fig:brunetti07}
\end{figure*}

Radio pointed observations of a complete sample of about 50 X--ray
luminous ($L_x \geq 5 \times 10^{44}$ erg s$^{-1}$) clusters at
redshift $z = 0.2 - 0.4$ have been recently carried out at 610 MHz
with the GMRT--{\it Giant Metrewave Radio Telescope} (Venturi et al.
2007, 2008). These observations were specifically designed to avoid
problems in the detection of the cluster-scale emission due to the
missing of short-baselines in the interferometric observations and
to image, at the same time, both compact and extended sources in the
selected clusters. Thus they allowed  a fair analysis of the
occurrence of Radio Halos in galaxy clusters. One of the most
relevant findings of these GMRT observations is that only a fraction
($\leq$ 30 \%) of the X-ray luminous clusters hosts Radio Halos
(Venturi et al. 2008); all Radio Halos being in 
merging--clusters. Remarkably, the fraction of clusters with Radio
Halos is also found to depend on the cluster X-ray luminosity (and
mass) decreasing at $\leq$ 10\% in the case of clusters at $z= 0.05
- 0.4$ with $L_X \approx 3\cdot 10^{44}-8\cdot 10^{44}$ erg s$^{-1}$
and suggesting the presence of some threshold in the mechanism for
the generation of these sources (Cassano et al. 2008). 
Although these studies demonstrate that no cluster-scale emission is detected
in the majority of cases, potentially synchrotron radio emission may be present
in all clusters at a level just below the sensitivity of radio
observations and this implies the importance to combine radio upper limits
and detections. 
Fig.\ref{fig:brunetti07} shows the distribution of
GMRT clusters in the radio power ($P_{1.4}$)-- cluster X-ray
luminosity ($L_x$) plane. The important point is that clusters with
similar $L_x$ (and redshift) have a bimodal distribution, with the
upper limits to the synchrotron luminosity of clusters with no hint
of Radio Halos that lie about one order of magnitude below the
region of the clusters with Radio Halos ($P_{1.4}$--$L_x$
correlation). It is probably useful to stress that these upper limits are
solid (conservative) being evaluated through injection of fake Radio
Halos into the observed datasets in order to account for the
sensitivity of the different observations to the cluster-scale
emission (Brunetti et al. 2007; Venturi et al. 2008).

Cluster bimodality and the connection of Radio Halos with cluster mergers,
suggest that Halos are {\it transient phenomena} that develop
in merging clusters.
From the lack of clusters in the region between Radio Halos
and upper limits, in the case that Halo-clusters evolve into radio-quiet 
clusters (and vice versa), the evolution should be fast, in a timescale
of $\approx$0.1-0.2 Gyr (Brunetti et al. 2007).
This observational picture suggests that turbulent re-acceleration of 
relativistic electrons may trigger the formation
of Radio Halos in merging clusters, in which case the acceleration
timescale of the emitting electrons is indeed $\approx$0.1-0.2 Gyr.
On the other hand, unless we admit the possibility of {\it ad hoc} fast
(on timescale of $\approx$0.1-0.2 Gyr) dissipation of the magnetic field
in clusters, the bimodal distribution in Fig.\ref{fig:brunetti07} cannot
be easily reconciled with {\it secondary} models.
In this case - indeed - Radio Halos should be common and some general
$P_{1.4}$--$L_x$ trend is predicted for all clusters
(e.g. Miniati et al. 2001; Dolag 2006; Pfrommer 2008; see discussion
in Brunetti et al. 2007 for more details).

\begin{figure*}[!t]\centering
\includegraphics[width=2\columnwidth]{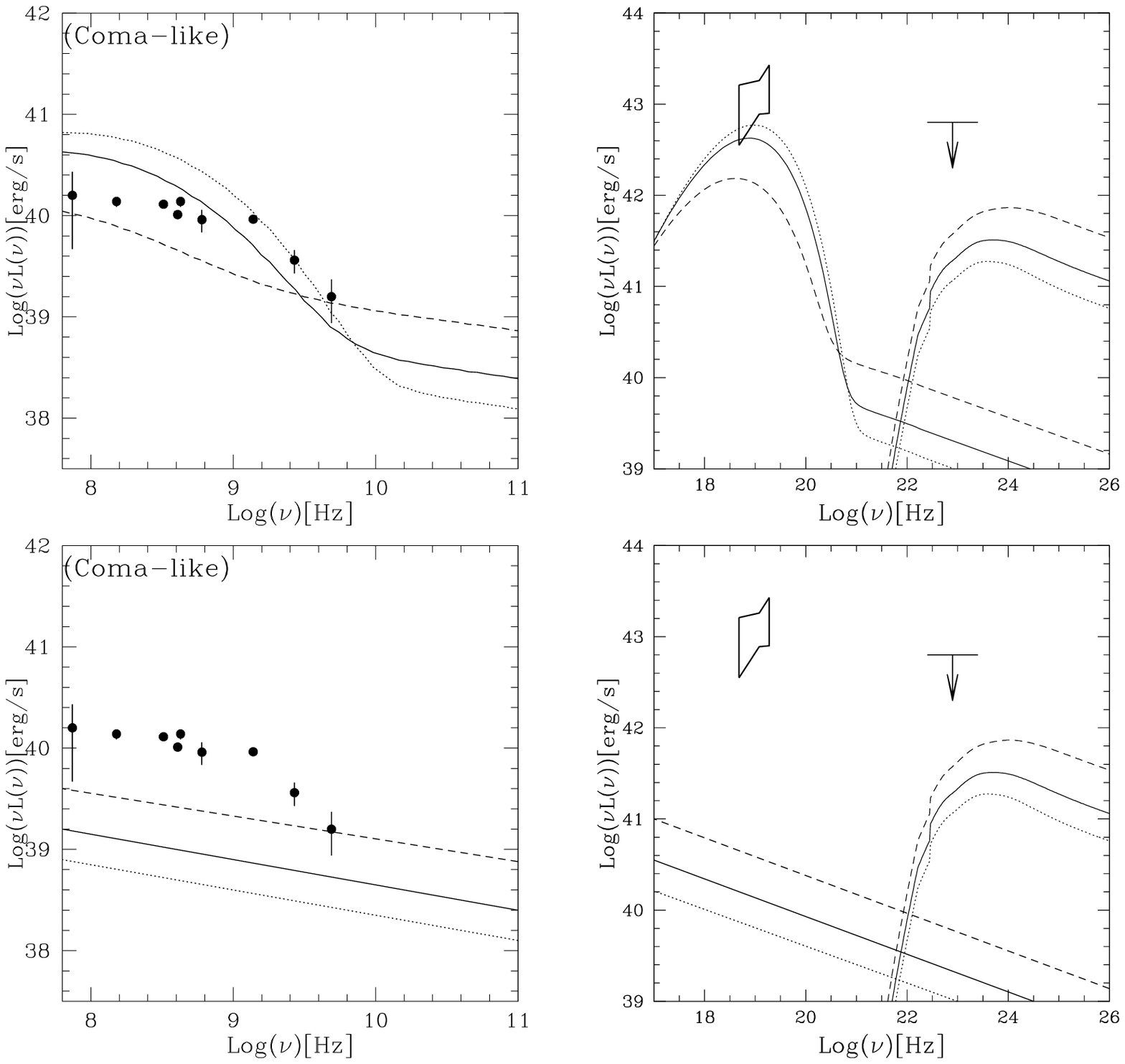}
\caption{Example of broad band spectrum for a Coma-like cluster.
{\bf Upper panels}: Synchrotron (left\footnote{SZ decrement at high
frequencies is not taken into account}) and IC and $\pi$-o emission
(right) calculated at t=0.75 Gyr from the injection of MHD
turbulence in the IGM. {\bf Lower panels}: Synchrotron (left) and IC
and $\pi$-o emission (right) calculated at t=1.75 Gyr from the
injection of MHD turbulence in the IGM; the energy density of
turbulence is set =0 for t$\geq$0.8 Gyrs. In all panels calculations
are shown assuming a ratio between the energy density of
relativistic and thermal protons = 3\% (dashed lines), 1\% (solid 
lines) and 0.5\% (dotted lines) at t=0 (with proton spectrum $\delta
=2.2$), a central cluster-magnetic field $B_o=1.5 \mu$G, a scaling
between field and thermal density $B(r) \propto n_{th}^{2/3}$,
$E_{CR}\propto E_{th}$, and a beta--model profile of the Coma
cluster. The energy density injected in Alfven modes between t=0-0.8
Gyr is $\approx$3\% of the thermal energy of the IGM. For the sake
of completeness we show radio data, BeppoSAX data and EGRET upper
limit for the Coma cluster, note - however - that model parameters
are not chosen to fit the data (see e.g. Brunetti \& Blasi 2005, for
model fittings to the Coma data).} \label{fig:broadspectrum}
\end{figure*}

\subsection{Limits on cosmic ray protons and future with the 
FERMI gamma-ray telescope}
\label{sec:CR}

Gamma ray observations with EGRET 
limit the energy of CR protons in a number of
nearby clusters to $\leq$20 \% of the energy of the IGM
(Reimer et al. 2003; Pfrommer \& Ensslin 2004).

The upper limits obtained for the clusters with no Radio Halos in
Fig.\ref{fig:brunetti07} imply that, regardless of the origin of the radio
emission, the synchrotron emission from secondary electrons
must be below the value of the upper limits. This allows to put a
corresponding limit to the energy density of the primary protons
from which secondaries are generated (Brunetti et al. 2007).

The limits are reported in Fig.\ref{fig:brunetti07}(right) as a function of the
magnetic field strength in the IGM and for different spectra of the primary
protons. By assuming $> \mu$G field, the upper limits to the
energy density of relativistic protons in the IGM are about one order
of magnitude more stringent than those obtained from EGRET upper limits  
(at least for relatively flat proton spectra). 
As a matter of fact, under these assumptions, no more than
1 \% of the energy density of the IGM in X--ray luminous clusters can
be in the form of relativistic protons; this would imply that even the
FERMI Gamma-ray telescope might not be sensitive enough to
detect the $\pi^o$--decay from these clusters.
On the other hand, for steeper spectra of relativistic protons, 
or lower values of the field, the synchrotron constraints
in Fig.\ref{fig:brunetti07}(right) become gradually less stringent
and the energy content of relativistic protons permitted by observations
is larger.
Since efficient particle acceleration mechanisms would
naturally produce flat proton spectra, it is unlikely that a population
of protons with steep spectrum stores a relevant fraction of the IGM 
energy and Fig.\ref{fig:brunetti07}(right) reasonably implies that
FERMI may detect the $\pi^o$ decay in a fairly large number
of clusters provided that the IGM is magnetised
at $\approx \mu$G level (or lower).
In this case, since $E_{CR}/E_{th}$ (and also the spectrum of relativistic
protons) can be constrained, the combination of gamma ray
measurements with deep radio upper limits in clusters without
Radio Halos will provide a novel tool to measure the magnetic
field strength in galaxy clusters.

\section{Re-acceleration models}
\label{sec:reacceleration}

\subsection{Basics}

The properties of the Mpc-scale
radio emission in galaxy clusters suggest that turbulence, generated
in cluster mergers, may play an important role in the acceleration of
the emitting particles.

The physics of collisionless turbulence and of stochastic particle
acceleration is complex and rather poorly understood, still several
calculations from {\it first principles} have shown that there is
room for efficient turbulent acceleration in the IGM.

One of the first points that have been realised is that
turbulent re-acceleration is not efficient enough to extract electrons
from the thermal IGM (e.g. Petrosian 2001) and thus a necessary assumption
in the re-acceleration scenario is that seed particles, relativistic
electrons with Lorentz factors
$\gamma \sim 100-500$, are already present in the IGM.
These seeds could be provided by the past activity of active galaxies
in the IGM or by the past merger history of the cluster (e.g.,
Brunetti et al. 2001), alternatively they could be secondary
electrons from p-p collisions (Brunetti \& Blasi 2005).

Particle re-acceleration in the IGM may be due to small scale Alfv\'en
modes (Ohno et al. 2002; Fujita et al. 2003; Brunetti et al. 2004) or
due to compressible (magnetosonic) modes (Cassano \& Brunetti 2005;
Brunetti \& Lazarian 2007).
In the case of Alfv\'enic models, the acceleration of relativistic
protons increases the damping of the modes and this severely limits
the acceleration of relativistic electrons when the energy budget of
relativistic protons is typically $\geq$ 5-7\% of that of the IGM
(Brunetti et al. 2004).
The injection process of Alfv\'en modes at small, resonant, scales in
the IGM and the assumption of isotropy of these modes represent the most
important sources of uncertainty in this class of models
(see discussion in Brunetti 2006 and Lazarian 2006b).
Alternatively, compressible (magnetosonic) modes might re-accelerate
fast particles via resonant Transit Time Damping (TTD) and non--resonant
turbulent compression (e.g., Cho et al. 2006; Brunetti \& Lazarian 2007).
The main source of uncertainty in this second class of re-acceleration
models is our ignorance on the viscosity in the IGM. Viscosity could indeed
severely damp compressible modes on large scales and inhibit particle
acceleration, although the super-Alfv\'enic (and sub-sonic)
nature of the turbulence in the IGM is expected to
suppress viscous dissipation (see discussion in Brunetti \& Lazarian 2007).

\subsection{The non-thermal spectrum of galaxy clusters}

The main goal of this Section is to suggest that the non-thermal
emission from clusters is a mixture of two main spectral components:
a long-living one that is emitted by secondary particles (and by
$\pi^o$ decay) continuously generated during p-p collisions in the
IGM, and a transient component that may be due to the re-acceleration of
relativistic particles by MHD turbulence generated (and then
dissipated) in cluster mergers. We model the re-acceleration of
relativistic particles by MHD turbulence in the most simple
situation in which only relativistic protons, accumulated during cluster
lifetime, are initially present in a turbulent IGM. These protons generate
secondary electrons via p-p collisions and in turns these secondaries (as
well as protons) are re-accelerated by MHD turbulence. More
specifically we adopt the Alfvenic model of Brunetti \& Blasi (2005)
and do not consider the possibility that {\it relic} primary
electrons in the IGM can be re-accelerated. Also we do not consider
the contribution to the cluster emission from fast electrons
accelerated at shock waves that develop during cluster mergers and
accretion of matter (see Ensslin, this conference).

An example of the expected broad band emission is reported in
Fig.\ref{fig:broadspectrum} for a Coma-like cluster
(curves in Figure are not fits to the data).

Upper panels show the non-thermal emission (synchrotron, IC, $\pi^o$ decay)
generated during a cluster merger, while the spectra in lower panels are
calculated after 1 Gyr from the time at which turbulence is dissipated, 
and thus they only rely with the long-living component of the non-thermal
cluster emission. As a first approximation the non thermal emission in the
lower panels does not depend on the dynamics of the clusters but 
only on the energy content
(and spectrum) of relativistic protons in the IGM (and on the magnetic
field in the case of the synchrotron radio emission).
On the other hand, the comparison  between the spectra in upper and lower
panels highlights the transient emission that is generated in connection
with the injection of turbulence during cluster mergers.

The results reported in Fig.\ref{fig:broadspectrum} have the potential 
to reproduce the radio bimodality observed in galaxy clusters
(Fig.\ref{fig:brunetti07}) :
Radio Halos develop in connection with
particle re-acceleration due to MHD turbulence in cluster mergers
where the cluster-synchrotron emission is considerably
boosted up (upper panel),
while a fainter long-living radio emission from secondary electrons
is expected to be common in clusters (lower panel); the level of this
latter component must be consistent with the radio upper
limits from radio observations (Fig.\ref{fig:brunetti07}, right).

An important point is that some level of gamma ray emission is expected.
However, we do not expect a direct correlation between
giant Radio Halos and the level of gamma ray emission from clusters,
indeed this level only depends on the content of protons
in the IGM (see Fig.\ref{fig:broadspectrum} caption).
FERMI is expected to shortly obtain crucial information on that.
By taking into account the constraints on the energy content of relativistic
protons obtained from the radio upper limits of clusters without Radio Halos
(Sect.2.3, Fig.\ref{fig:brunetti07}) the gamma ray emission
from a Coma--like cluster (assuming $E_{CR} \propto E_{th}$ and
B$\approx \mu$G as in Fig.\ref{fig:broadspectrum}) is expected
$\approx$10-20 times below present EGRET upper limits and
no substantial amplification of this signal is expected
in clusters with Radio Halo.
On the other hand, in the case of smaller magnetic fields
in the IGM, the gamma ray luminosity
of clusters can be larger because a larger proton-content is permitted
by radio upper limits
(a more detailed discussion on the gamma ray properties of clusters
in the re-acceleration scenario is given in Brunetti et al. in prep.).

Fig.\ref{fig:broadspectrum} shows that a direct correlation is expected
between Radio Halos and IC emission in the hard X-rays since
the two spectral components are emitted by (essentially) the same population
of relativistic electrons. Because the ratio between IC and radio luminosity
depends on the magnetic field in the IGM, future hard X-ray telescopes
(Simbol-X, NuSTAR and Next) are expected to
detect a fairly large number of clusters with Radio Halos in the case
that the IGM is magnetised at $\approx 1 \mu$G level (or lower).

\subsection{Low frequency radio emission from galaxy clusters}
\label{sec:radiolow}

\begin{figure}[!t]
\includegraphics[width=\columnwidth]{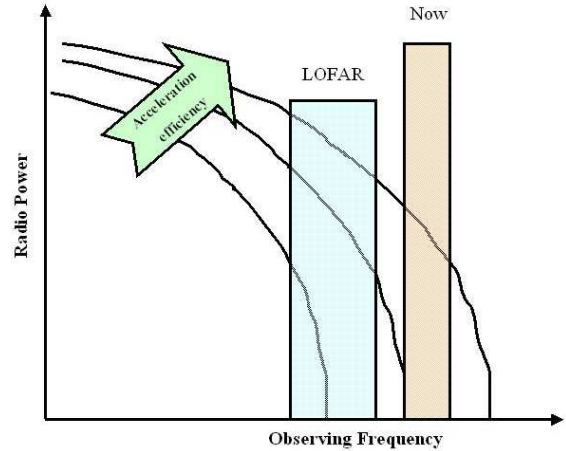}%
\caption{Picture showing the connection between acceleration
efficiency and spectrum of Radio Halos in the re-acceleration scenario.
Boxes indicate the frequency range in which Radio Halos can be studied
with present radio telescopes and with LOFAR.}
\label{fig:schema}
\end{figure}

The maximum energy at which electrons can be re-accelerated by turbulence
in the IGM, and ultimately the frequencies at which the spectra
of Radio Halos cut off (e.g. Fig.\ref{fig:broadspectrum}), depend on
the acceleration efficiency that essentially
increases with the level of turbulence in the IGM.
The spectral cut off affects our ability to detect Radio Halos in
the universe, introducing a strong bias against observing them at
frequencies substantially larger than the cut off frequency.
In the context of the re-acceleration scenario, presently known
Radio Halos must result from the rare, most energetic merging events
and therefore be hosted only in the most massive and hot
clusters (Cassano \& Brunetti 2005).
On the other hand, it has been realised that
a large number of Radio Halos should be formed during much more common
but less energetic mergers and should be visible
only at lower frequencies because of their ultra steep spectral
slope (Fig. \ref{fig:schema}).
Thus the fraction of clusters with Radio Halos increases at lower
observing frequencies and LOFAR and LWA, that will observe galaxy
clusters with unprecedented sensitivity at low radio frequencies,
have the potential to catch the bulk of these sources
in the universe (Cassano et al. 2006).

This is a unique expectation of the re-acceleration scenario.
In particular, following Cassano et al. (2008), it has been calculated that
observations at 150 MHz
have the potential to increase the number of giant Radio Halos observed
in galaxy clusters with mass $M\leq 10^{15}\,M_{\odot}$ by almost one
order
of magnitude with respect to observations at 1.4 GHz, while this increase
is expected to be smaller for clusters with larger mass.

\section{Summary}
\label{sec:summary}

Recent observations show that Radio Halos are not common in galaxy clusters
suggesting that they are transient sources in cluster mergers.
These facts support
the idea that MHD turbulence generated in cluster mergers may
play an important role in the re-acceleration of particles in the IGM.

We have suggested that in the context of the re-acceleration scenario the
non-thermal emission from galaxy clusters is a combination
of two components : a long-living one that is emitted
by secondary particles (and by $\pi^o$ decay) continuously generated
during p-p collisions in the IGM, and a transient component due to
the re-acceleration of relativistic particles by MHD turbulence
generated (and then dissipated) in cluster mergers.

The latter component may naturally explain Radio Halos in merging and massive
clusters and their statistical properties, on the other hand FERMI
may detect the long-living gamma ray
emission from clusters due to the decay of $\pi^o$ produced via
collisions between relativistic and thermal protons.
In particular, given the constraints on the relativistic protons obtained
from upper limits to the synchrotron emission due to secondaries
in X-ray luminous clusters, we conclude that the detection of a fairly
large number of clusters with FERMI would imply that the magnetic field
in the IGM is at $\approx \mu$G level (or lower).
In this case, the pool of future hard X-ray detectors (NuSTAR,
Simbol-X and Next) should detect IC emission from a fairly large number
of merging clusters with Radio Halos.

An important expectation of the re-acceleration scenario is the
presence of a population
of Radio Halos in the universe that should emerge only at low
radio frequencies, and this can be easily tested in a few years
with LOFAR and LWA.
The detection of Radio Halos with ultra-steep spectral slope will thus
provide compelling support to particle re-acceleration due to turbulence
in the IGM.

\noindent
{\it Acknowledgment}: The author is very grateful to the organizing
committee of the Conference and also thanks P.Blasi, R.Cassano, V.Dogiel,
S.Gabici, A.Lazarian and F.Vazza for the ongoing theoretical collaborations.
This work is partially supported by PRIN-INAF2008 and ASI-INAF I/088/06/0.

\end{document}